\documentclass[aps,prl,twocolumn]{revtex4}
\usepackage{graphicx}
\usepackage{amsmath}
\usepackage{amssymb}
\usepackage{array}
\usepackage{mathptmx}
\usepackage{anyfontsize}
\usepackage{t1enc}
\usepackage[colorlinks,linkcolor={blue},citecolor={blue},urlcolor={blue}]{hyperref}
\begin{document}

\title{A simple model for dynamic heterogeneity in glass-forming liquids}
\author{Rajib K Pandit\footnote{Present address: Exact Sciences,
    Redwood City, 94063 CA, USA }} 
\author{Horacio E. Castillo} 
\email[]{castillh@ohio.edu}
\affiliation{
  Department of Physics and Astronomy and Nanoscale
  and Quantum Phenomena Institute, Ohio University, Athens, Ohio 45701,
  USA}

\date{\today}

\begin{abstract}
  Liquids near the glass transition exhibit dynamical heterogeneity,
  i.e.~local relaxation rates fluctuate strongly over space and
  time. Here we introduce a simple continuum model that allows for
  quantitative predictions for the correlators describing these
  fluctuations. We find remarkable agreement of the model predictions
  for the dynamic susceptibility $\chi_4(t)$ with numerical results
  for a binary hard-sphere liquid and for a Kob-Andersen Lennard-Jones
  mixture. Under this model, the lifetime $\tau_{\rm ex}$ of the
  heterogeneities has little effect on the position $t = t_4 \sim
  \tau_{\alpha}$ of the peak of $\chi_4(t)$, but it controls the decay
  of $\chi_4(t)$ after the peak, and we show how to estimate it from
  this decay.
\end{abstract}


\maketitle

\paragraph{Introduction.}
When cooled or compressed fast enough, most liquids will undergo a
glass transition, where the characteristic timescale for relaxation,
the $\alpha$-relaxation time
$\tau_{\alpha}$~\cite{Donth2001,Angell2000,Debenedetti2001} grows
smoothly but extremely rapidly. Together with the increase in
relaxation time, {\em dynamical heterogeneity\/} emerges: relaxation
becomes much slower in some regions than in
others~\cite{Angell2000,Ediger2000,Berthier2011a,Sillescu1999}.
In this work we introduce a phenomenological model for dynamic
heterogeneity which is directly based on this intuitive description
and uses a local relaxation rate $\gamma(\vec{r},t) \equiv
1/\tau_{\vec{r}}(t)$ as its basic variable. The model allows to write
the multi-point correlators measured in numerical simulations in terms
of the two-point correlator $s(\vec{q},t,t')$ of $\gamma(\vec{r},t)$,
which has a direct interpretation in terms of the size and lifetime of
the heterogeneous regions. These quantities can provide a more direct
connection
between numerical simulation results and the results of
experiments~\cite{Ediger2000,Sillescu1999,Richert2015,Paeng2015} on
dynamical heterogeneity. More immediately, we show in this work that:
(i) numerical simulation data for two glass-forming
systems~\cite{Flenner2011,Flenner2014} are consistent with predictions
from the model; (ii) the model explains quantitatively and in a
transparent way some of the general features~\cite{Lacevic2003} of the
numerical results, and (iii) the model provides a new, simpler way to
extract an estimate of the lifetime of the heterogeneities from
numerical simulation data.

\paragraph{Probing relaxation in numerical simulations}
requires detecting changes in individual particle positions or
orientations over a certain time interval of length $t$,
i.e.~measuring a two-time observable $C(t)$~\cite{Berthier2011a}. For
example the overlap $C(t) = F_o(t)$~\cite{Flenner2011} gives the
fraction of ``slow particles'', i.e.~those particles that over the
interval $t$ have had displacements shorter than a certain distance
$a$. Because of dynamic heterogeneity, the fraction of slow particles
varies over space and time. This variation can be probed by measuring
four-point functions such as the {\em four-point structure factor\/}
$S_4(\vec{q},t)$~\cite{Dasgupta1991,Lacevic2003,Berthier2011}, which
is the Fourier-space correlation function of the {\em local\/} density
of slow particles, or, equivalently, the structure factor of the slow
particles, (up to a trivial additive constant). The limit $\chi_4(t) =
\lim_{q \to 0}
S_4(\vec{q},t)$~\cite{Lacevic2003,Toninelli2005,Parsaeian2008,Flenner2011,Berthier2011},
called the {\em dynamic susceptibility\/}, captures contributions from
dynamic heterogeneities of all spatial extents, and has been a central
quantity in numerical studies of dynamical heterogeneity, where it has
been often used to provide an overall measure of the degree of
heterogeneity, or, in other words, of the intensity of the
fluctuations.  Although there have been some efforts to predict or
interpret the time dependence of the four-point dynamic susceptibility
$\chi_4(t)$, particularly for times up to the $\alpha$-relaxation
time~\cite{Toninelli2005}, much less is known about it for times $t >
\tau_{\alpha}$, and about how the time evolution of $\chi_4(t)$ is
connected with the time evolution of the heterogeneous regions.

\paragraph{Observables and data.}  
For simplicity, we consider only systems, such as supercooled liquids,
where (i) the heterogeneity is dynamic, not static, and therefore all
thermodynamic averages are translation invariant, and (ii) the
dynamics is time-translation-invariant (TTI), e.g. aging is not
present. We also restrict ourselves to observables involving only one
time interval from time $0$ to time $t$.  We probe the dynamics by
using the microscopic overlap function
$w_n(t)=\theta[a-|\vec{r}_n(t)-\vec{r}_n(0)|]$, where $\theta(x)$ is
the Heaviside step function, $\vec{r}_n(t), n=1,\cdots,N$ is the
position of the $n$-$th$ particle at time $t$, and $a$ is a
characteristic distance that is larger than the typical amplitude of
vibrational motion. We introduce the local
relaxation function $C_{\vec{r}}(t)$,
\begin{equation}
  C_{\vec{r}}(t) \equiv \rho^{-1} 
  \sum_{n=1}^N w_n(t)
  \delta(\vec{r}_n(0)-\vec{r}), \; \text{with} \; \langle C_{\vec{r}}(t)
  \rangle = C(t),
\end{equation}
where $\rho$ is the average particle density, $\langle \cdots \rangle$
is an average over thermal fluctuations, and $C(t)$ is the global
two-point correlator, i.e. the average overlap $C(t) = F_o(t) \equiv
N^{-1} \sum_{n=1}^N \langle w_n(t)
\rangle$~\cite{Flenner2011}. Fluctuations are characterized by the
four-point dynamic structure factor
$S_4(\vec{q},t)$~\cite{Dasgupta1991,Lacevic2003},
\begin{eqnarray}
  \label{eq:S4-def}
  && \hspace{-0.75cm} S_4(\vec{q},t) \equiv 
  {\rho^2}{N^{-1}}
  \!\!\! \textstyle\int \! d^d \! r
  \, {\rm e}^{-i \vec{q}\cdot{\vec{r}}}
  \langle [C_{\vec{r}}(t)-C(t)]
    [C_{\vec{0}}(t)-C(t)] \rangle \\
  \label{eq:S-4}
  && \hspace{-0.4cm} = N^{-1}
  \!\!\! \sum_{n,n^{\prime}=1}^{N} \! \!
  \Big\langle w_n(t) w_{n^\prime}(t)
  e^{i\vec{q}\cdot( \vec{r}_n(0) -  \vec{r}_{n^\prime}(0) )} \Big\rangle
  -\delta_{\vec{q},0}N C^2(t). 
\end{eqnarray}
Numerical results~\cite{note-details-supplemental,
  OhioSupercomputerCenter1987}
are discussed for a
3D hard-sphere binary mixture (HARD)~\cite{Flenner2011}, and for a 3D
Kob-Andersen Lennard-Jones binary mixture
(KALJ)~\cite{Flenner2014}.

\paragraph{Continuum model for dynamic heterogeneity.}
The model we introduce focuses on the $\alpha$-relaxation regime, 
i.e.~times $t$ during the second step in the two-step relaxation. It is
based on two assumptions, which we discuss below.

\paragraph{Motivation for Assumption 1} 
Even though four-point functions were introduced to describe the
collective phenomenon of dynamic heterogeneity, they capture much more
than that. In Ref.~\cite{Pandit2022} it was shown that at very long
times $t$, the collective contributions to the four-point function
$S_4(\vec{q},t)$ are negligible compared to the single-particle,
spatially uncorrelated, $\vec{q}$-independent contribution $S_4^{\rm
  sp}(\vec{q},t) = \chi_4^{\rm sp}(t) \approx C(t) -
C^2(t)$. 
To reproduce this behavior, we introduce:

\paragraph{Assumption 1} {\em The local relaxation function
  $C_{\vec{r}}(t)$\/} is the sum of two mutually 
independent random variables, the collective part $C^{\rm
  coll}_{\vec{r}}(t)$ and the single particle part $C^{\rm
  sp}_{\vec{r}}(t)$,
\begin{eqnarray}
  && C_{\vec{r}}(t) = C^{\rm coll}_{\vec{r}}(t)
  + C^{\rm sp}_{\vec{r}}(t), \quad \text{with} \quad
  \label{eq:Cr-Ccoll-Csp} \\
  &&
  \langle C^{\rm coll}_{\vec{r}}(t) \rangle = C(t)
  \quad \text{and}
  \quad \langle C^{\rm sp}_{\vec{r}}(t) \rangle = 0. 
  \label{eq:Cr-sp-avg-0}
\end{eqnarray}
We interpret all correlations between different particles as
``collective'', thus the single particle part $C^{\rm
  sp}_{\vec{r}}(t)$ is spatially uncorrelated at equal times,
\begin{equation}
  \hspace{-0.3cm}
  G_4^{\rm sp}(\vec{r},t)
  \equiv \rho \langle C^{\rm sp}_{\vec{r}}(t) C^{\rm sp}_{\vec{0}}(t) \rangle
  \propto \delta(\vec{r}), 
  \;
  \text{and} \;
  {S}_4^{\rm sp}(\vec{q},t) 
  = \chi_4^{\rm sp}(t).
\label{eq:S4sp-tdiag-const}
\end{equation}

\begin{figure}[ht!]
  \centering
  \hspace{-0.8cm}
  \includegraphics[width=\columnwidth]{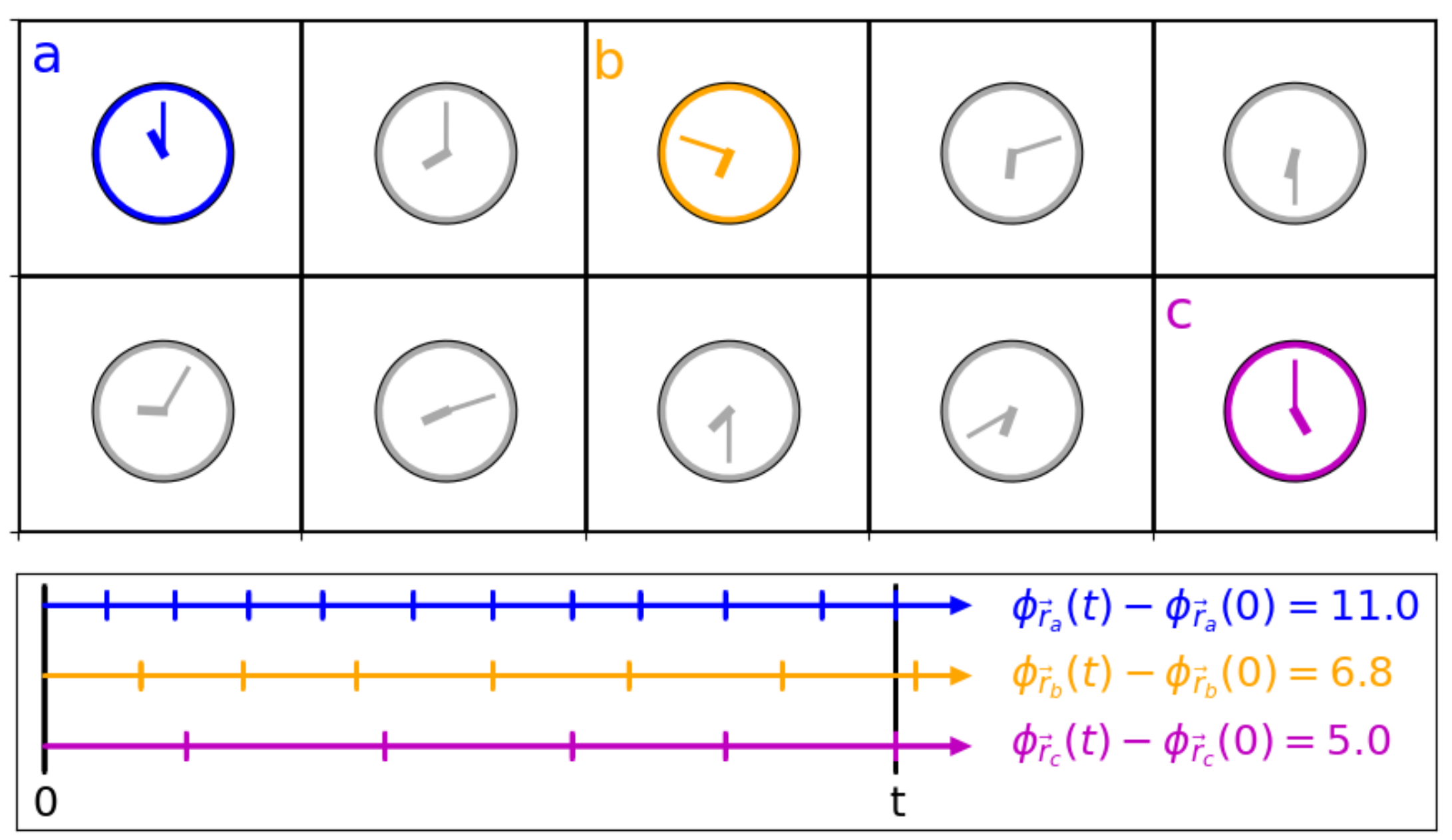}
  \caption{{\em Converting the idea of space and time dependent
      relaxation times $\tau_{\vec{r}}(t)$ into the idea of a ``local
      clock''.}
    For each region in the system there is an individual local
    clock $\phi_{\vec{r}}(t)$ (Eq.~\ref{eq:phir_local_clock}) that
    counts how many local relaxation times have elapsed up to time
    $t$. Each tic mark on a horizontal line represents one relaxation
    time. For example, of the three highlighted regions, $a$, $b$ and
    $c$, region $c$ has relaxed the slowest, with $\Delta
    \phi_{\vec{r}_c}(t) = \phi_{\vec{r}_c}(t)-\phi_{\vec{r}_c}(0) =
    5.0$ relaxation times elapsed between time $0$ and time
    $t$. Regions $a$ and $b$ have relaxed generally faster, so the
    corresponding numbers of elapsed relaxation times between times
    $0$ and $t$ are larger, in this case $\Delta \phi_{\vec{r}_b}(t) =
    6.8$ and $\Delta \phi_{\vec{r}_a}(t) = 11.0$.
  }\label{fig:local_clock}
\end{figure}
%
\paragraph{Motivation for Assumption 2} 
Dynamic heterogeneity has been described as involving relaxation times
differing in different spatial
regions~\cite{Sillescu1999,Ediger2000,Richert2015}, i.e. some regions
being ``fast'' and others being ``slow''. Our model
directly translates this intuitive description into quantitative
predictions by defining a ``local clock'' $\phi_{\vec{r}}(t)$
(Fig.~\ref{fig:local_clock}), which instead of counting in units of
seconds, counts in units of the local relaxation time
$\tau_{\vec{r}}(t)$,
\begin{equation}
  \phi_{\vec{r}}(t) \equiv \int^t dt'/\tau_{\vec{r}}(t').
  \label{eq:phir_local_clock}
\end{equation}
Thus $\Delta \phi_{\vec{r}}(t) \equiv
\phi_{\vec{r}}(t)-\phi_{\vec{r}}(0)$ represents the number of
relaxation times elapsed between times $0$ and $t$
in the region around $\vec{r}$
(Fig.~\ref{fig:local_clock}). Naively one could expect $C^{\rm
  coll}_{\vec{r}}(t)$ to depend only on $\Delta \phi_{\vec{r}}(t)$,
i.e. $C^{\rm coll}_{\vec{r}}(t) = {\mathcal C} \left[\Delta
  \phi_{\vec{r}}(t)
  \right]$~\cite{Mavimbela2012,Mavimbela2019,Pandit2019}. Here
${\mathcal C}(x)$ doesn't fluctuate, it is a fixed monotonous
decreasing function, with ${\mathcal C}(1) = {\rm e}^{-1} \: {\mathcal
  C}(0)$, that represents the shape of the local relaxation function,
for example ${\mathcal C}(x) = f_0 \exp(-x)$ for simple exponential
relaxation. This {\em ansatz}, however, does not reproduce
$S_4(\vec{q},t)$, because spatial density fluctuations in the initial
state of the system provide two $\vec{q}$-dependent
contributions~\cite{Pandit2022} to $S_4(\vec{q},t)$. One is $S_4^{\rm
  st}(\vec{q},t) = C^2(t) S(\vec{q})$ (for $q \ne 0$), where
$S(\vec{q})$ is the static structure factor; the other, $S_4^{\rm
  mc}$~\cite{note-details-supplemental}
was neglected in the case of $\chi_4(t) = \lim_{q \to 0}
S_4(\vec{q},t)$. This motivates multiplying ${\mathcal C} \left[
  \Delta \phi_{\vec{r}}(t) \right]$ by the initial local
particle density $\rho(\vec{r},0)$, which leads to:

\paragraph{Assumption 2} {\em The collective contribution\/} is
\begin{eqnarray}
  && \quad
  C^{\rm coll}_{\vec{r}}(t) = \rho^{-1} \rho(\vec{r},0) \;
  {\mathcal C} \left[\phi_{\vec{r}}(t)-\phi_{\vec{r}}(0) \right], 
  \label{eq:Cr-calC-phi-r} \\
  &&
  \text{where} \quad 
  \label{eq:gamma-def-phir-quick}
  \partial\phi_{\vec{r}}(t)/\partial t \equiv 1/\tau_{\vec{r}}(t) = 
  \gamma({\vec{r}},t) 
\end{eqnarray}
is the local relaxation rate. 
For simplicity, we assume that $\rho(\vec{r},t_1)$ and
$\phi_{\vec{r'}}(t_2)$ are mutually independent random variables, for
arbitrary positions $\vec{r}$, $\vec{r'}$, and times $t_1, t_2$, and
that they are smooth and slowly varying in space and time.

\paragraph{Motivation for the definition of $\tau_{\rm ex}$} 
The exchange time $\tau_{\rm ex}$ has been defined as the
characteristic time for a ``fast'' region becoming ``slow'' or
viceversa~\cite{Sillescu1999,Ediger2000,Richert2015}. This translates
directly into defining it as the characteristic
time for the autocorrelation function of the relaxation rate
$\gamma({\vec{r}},t) = 1/\tau_{\vec{r}}(t)$.

\paragraph{Definition of the exchange time $\tau_{\rm ex}$}
\label{sec:assumption-tau-ex}
\begin{eqnarray}
  && \hspace{-0.3cm} 
  \tau_{\rm ex}\equiv
  \frac{\int_{0}^{\infty} t \, \chi_2^{\phi}(t) \, dt}
       {\int_{0}^{\infty} \chi_2^{\phi}(t) \, dt},
  \label{eq:t-ex-def-chi2e}
  \quad \text{where} \quad
  \chi_2^{\phi}(t) \equiv
  \lim_{q \to 0} \; s({\vec{q}},t), \quad
  \\
  && 
  \text{and} \;\;
  s({\vec{q}},t) \equiv
  V^{-1} \!\! \textstyle\int d^d\!r \,
  {\rm e}^{-i \vec{q}\cdot{\vec{r}}}
  \langle \delta \gamma({\vec{r}},t)
  \delta \gamma({\vec{0}},0) \rangle. \quad  
  \label{eq:def-s}
\end{eqnarray}
Here $\chi_2^{\phi}(t)$ and $s({\vec{q}},t)$ are two-point functions
of the fluctuations $\delta\gamma(\vec{r},t) \equiv
\gamma(\vec{r},t) - \langle \gamma(\vec{r},t) \rangle$ of the
relaxation rate. The limit $q \to 0$ in Eq.~(\ref{eq:t-ex-def-chi2e})
includes fluctuations at all spatial lengthscales, as in the
definition of
$\chi_4(t)$~\cite{Lacevic2003,Toninelli2005,Parsaeian2008,Flenner2011,Berthier2011}.

\begin{figure}[ht!]
  \centering
  \includegraphics[width=\columnwidth]{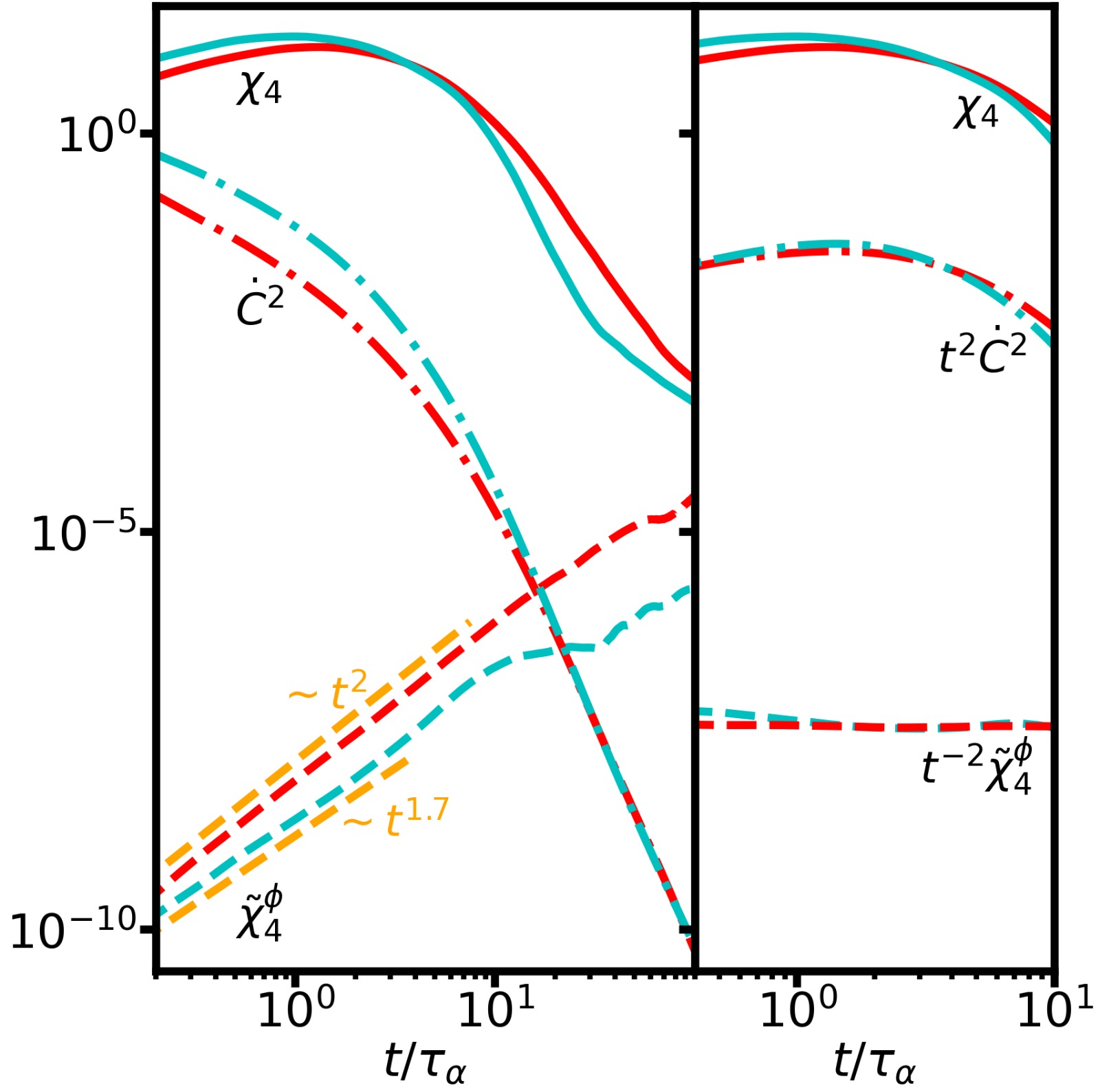}
  \caption{Numerical results~\cite{OhioSupercomputerCenter1987}
    for KALJ (cyan) and HARD (red).
    Left panel: $\chi_4(t)$ for KALJ for $T=0.50$ and for
    HARD at $\varphi=0.58$, $10^{6} \dot{C}^2(t)$ for KALJ at
    $T=0.50$ and $10^{10} \dot{C}^2(t)$ for HARD at $\varphi=0.58$,
    $10^{-21} \tilde{\chi}_{4}^{\phi}(t)$~\cite{note-tilde} for HARD at
    $\varphi=0.58$, and $10^{-17} \tilde{\chi}_{4}^{\phi}(t)$ for KALJ
    at $T=0.50$. $\tilde{\chi}^{\phi}_4(t)$ is extracted from the data
    by $\tilde{\chi}^{\phi}_4(t) = [\chi_4(t) - \chi_{4,b}^{(0)}(t)]
    \dot{C}^{-2}(t)$. The yellow dashed lines represent a $\propto
    t^2$ time dependence (frozen heterogeneity approximation for
    $\tilde{\chi}_{4}^{\phi}(t)$), and a $\propto t^{1.7}$ time
    dependence.
    %
    %
    %
    %
    %
    %
    Right panel: $\chi_4(t)$ for KALJ at $T=0.50$ and for HARD at
    $\varphi=0.58$, $t^2 \dot{C}^2(t)$ for KALJ at $T=0.50$ and for
    HARD at $\varphi=0.58$, $10^{-10} t^{-2}
    \tilde{\chi}_{4}^{\phi}(t)$ for HARD at $\varphi=0.58$ and for
    KALJ at $T=0.50$.  }\label{fig:chi-phi4}
\end{figure}

\paragraph{Results: Prediction for the dynamic susceptibility $\chi_4(t)$}
By Taylor expanding Eq.~(\ref{eq:Cr-calC-phi-r}) up to quadratic order in 
$\delta \gamma$, Assumptions $1$ and $2$ lead to
\begin{eqnarray}
   && \hspace{-1.0cm} {\mathcal C}'(t/\tau)
   \approx \tau \dot{C}(t), \; \text{where} \; C(\tau) = {\rm e}^{-1}, 
   \; \text{and} 
   \label{eq:CalCprime-taudotC}
   \\ 
   && \hspace{-1.0cm}
   \chi_4(t) \equiv \lim_{q \to 0} S_{4}({\vec{q}},t)  = 
   \dot{C}^2(t) \; \tilde{\chi}^{\phi}_4(t)
   + \chi_{4,b}^{(0)}(t)
   , \; \text{with}
   \label{eq:chi4-chi4tilde-chi40}
   \\
   && \hspace{-1.0cm}
   \chi_{4,b}^{(0)}(t) 
   \equiv 
   C(t) + [N^{-1} \langle (\delta N)^2 \rangle - 1] C^2(t)
   \label{eq:chi4b0-def}
\end{eqnarray}
representing a background due to single particle and initial density
fluctuations. 

Eq.~(\ref{eq:chi4-chi4tilde-chi40}) is the main result of this
work.
The factor $\dot{C}^2(t)$ comes directly from the linear term in the
Taylor expansion, and its presence is unavoidable for any {\em
  ansatz\/} that contains a factor like ${\mathcal C}
\left[\phi_{\vec{r}}(t)-\phi_{\vec{r}}(0) \right]$ that attempts to
represent the effects of fluctuating local relaxation
rates.
The factor $\tilde{\chi}^{\phi}_4(t)$~\cite{note-tilde} is predicted
to be a positive 
monotonously increasing function of $t$, which has no special feature
for $t \sim \tau_{\alpha}$ (Fig.~\ref{fig:chi-phi4}). 
Initially, $\chi_4(t)$ grows rapidly due to the rapid growth of
$\tilde{\chi}^{\phi}_4(t)$, but this growth gets cut off by the sharp
decrease of $\dot{C}^2(t)$ near $t \sim \tau_{\alpha}$, corresponding
to the fact that $\tau_{\alpha}$ is the characteristic time for the
decay of $C(t)$. This leads to
$\chi_4(t)$ having its peak for $t \sim
\tau_{\alpha}$~\cite{Lacevic2003,Toninelli2005}, even if the exchange
time is much longer~\cite{Kim2013}. Another way of looking at this is
to rewrite $\dot{C}^2(t) \tilde{\chi}^{\phi}_4(t) = [t^2 \dot{C}^2(t)]
[t^{-2} \tilde{\chi}^{\phi}_4(t)]$, where $t^{-2}
\tilde{\chi}^{\phi}_4(t)$ is still featureless for $t \sim
\tau_{\alpha}$ (Fig.~\ref{fig:chi-phi4}), but $t^2 \dot{C}^2(t)$ has a
peak at $t \sim \tau_{\alpha}$, which leads to the peak in
$\chi_4(t)$. $\tilde{\chi}^{\phi}_4(t)$ controls the height of the
peak of $\chi_4(t)$, and provides an overall measure of the
heterogeneity, because it is approximately proportional to the mean
quadratic drift $\langle [\Delta \phi_{\vec{r}}(t) - \langle \Delta
  \phi_{\vec{r}}(t)\rangle ]^2 \rangle = \langle [\int_0^t \delta
  \gamma(\vec{r},t) ]^2 \rangle = \int_{0}^{t} \! dt'' \!\!
\int_{0}^{t} \! dt' {\chi}_2^{\phi}(t''-t')$ of the local clocks,
\begin{eqnarray}
   && \hspace{-1.0cm}
   \tilde{\chi}^{\phi}_4(t) \equiv 
   \rho \tau^2
   \!\! \textstyle\int_{0}^{t} \! dt'' \!\!
   \textstyle\int_{0}^{t} \! dt'
   \tilde{\chi}_2^{\phi}(t''-t'),
   \quad \text{with}
   \label{eq:tilde-chi4-phi-def}
   \\
   && \hspace{-1.0cm}
   \tilde{\chi}_2^{\phi}(t) \equiv
   {\chi}_2^{\phi}(t)
   +      
   \lim_{q \to 0}
   \rho^{-1} 
   \left\{[S_c(\cdot)-1]*s(\cdot,t)\right\}(\vec{q}), 
   \label{eq:tilde-chi2-def}
\end{eqnarray}
where $S_c({\vec{q}}) \equiv [S({\vec{q}}) - \rho \delta_{\vec{q},
    \vec{0}}]$ is the connected part of the static structure factor
and $f*g$ denotes a convolution. In fact, in most cases,
$\tilde{\chi}_2^{\phi}(t) \approx
{\chi}_2^{\phi}(t)$~\cite{note-tilde}.

\begin{figure}[ht!]
  \centering
  \hspace{-0.8cm}
  \includegraphics[width=1.065\columnwidth]{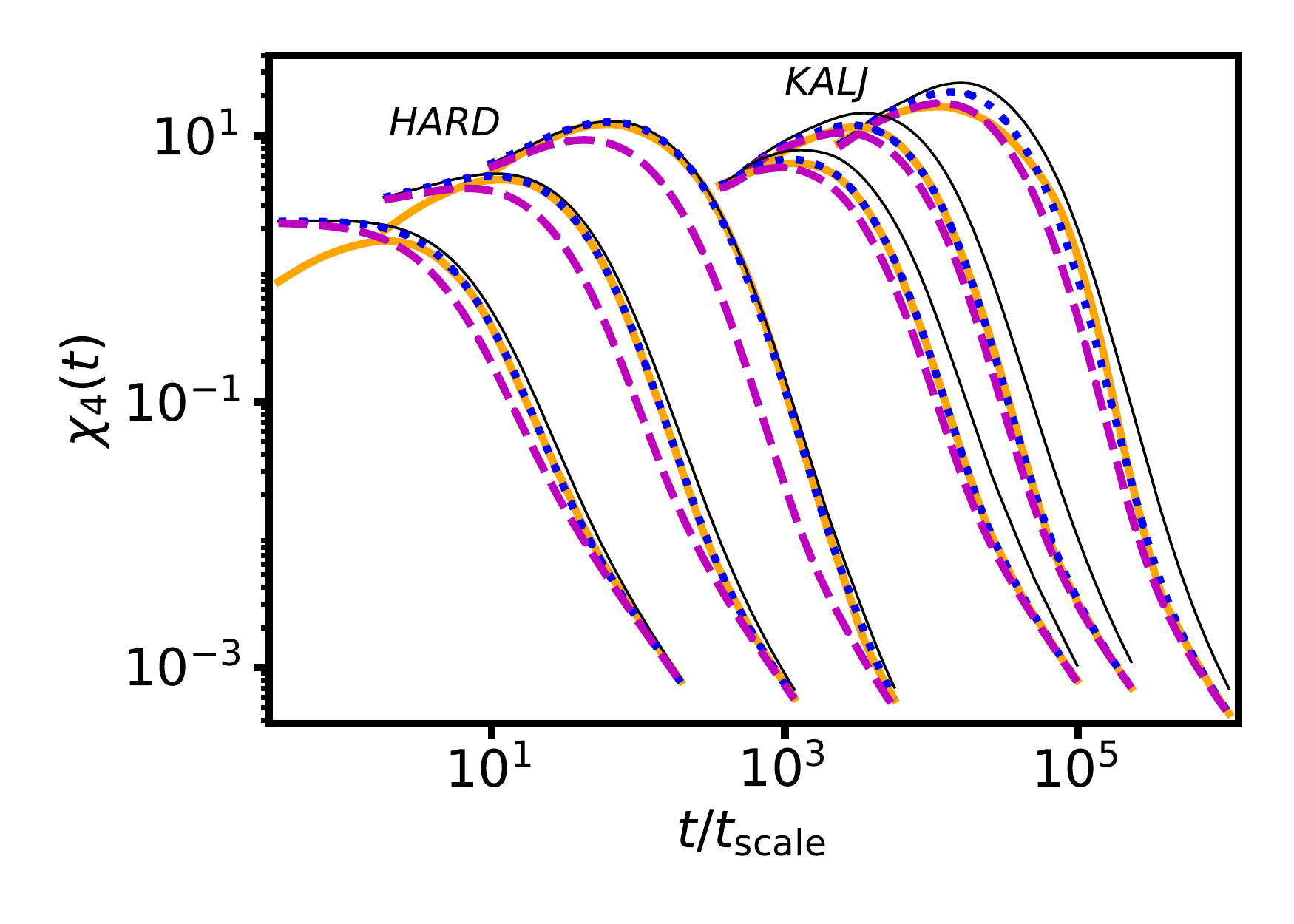}
  \caption{Comparison of our model's predictions for $\chi_4(t)$,
    Eqs.~(\ref{eq:chi4-chi4tilde-chi40})-(\ref{eq:tilde-chi4-phi-def}),
    with numerical simulation
    results~\cite{OhioSupercomputerCenter1987}
    (orange full lines) for: HARD
    for packing fractions $\varphi=0.55, 0.57, 0.58$ (left side, left
    to right) and KALJ for temperatures $T=0.60, 0.55, 0.50$ (right
    side, left to right). $t_{\rm scale}$ is $1$ for HARD and $0.005$
    for KALJ. Model predictions are shown for: (i) the frozen
    heterogeneity approximation $\chi_4(t) = \chi_{4, \rm fh}(t)$
    (black thin lines); (ii)
    $\tilde{\chi}_2^{\phi}(t) = \tilde{\chi}_2^{\phi}(0) \, {\rm
      e}^{-t/\tilde{\tau}_{\rm ex}}$ with $\tilde{\tau}_{\rm ex} =
    \tau_{\alpha}$~\cite{note-tilde} (magenta dashed
    lines); and (iii) $\tilde{\chi}_2^{\phi}(t) =
    \tilde{\chi}_2^{\phi}(0) \, {\rm e}^{-t/\tilde{\tau}_{\rm ex}}$
    with $\tilde{\tau}_{\rm ex}$ chosen by fitting the data with
    Eq.~(\ref{eq:tilde-chi4-exp-case}) (blue dotted lines).
  }\label{fig:compare-model-simulations}
\end{figure}

\paragraph{Results: Upper bound for $\chi_4(t)$ and frozen
  heterogeneity approximation.}
Eqs.~(\ref{eq:chi4-chi4tilde-chi40}) and (\ref{eq:tilde-chi4-phi-def})
imply there is an upper bound $\chi_{4, \rm fh}(t)$ for
$\chi_4(t)$, 
\begin{eqnarray}
   && \hspace{-1.0cm}
   \chi_4(t) \le \chi_{4, \rm fh}(t) \equiv
   \tilde{a}_{\phi} t^2 \dot{C}^2(t) + \chi_{4,b}^{(0)}(t), \quad \text{with}
   \label{eq:chi4-ineq-a-phi-chi4b0}
   \\
   && \hspace{-1.0cm}
   \tilde{a}_{\phi} \equiv
   \rho \tau^2 \tilde{\chi}_{2, M} \quad
   \text{and} \quad
   \tilde{\chi}_{2, M} \equiv \sup_{t}
   \tilde{\chi}_2^{\phi}(t) < \infty, 
   \label{eq:tilde-chi-2M-def}
\end{eqnarray}
where the parameter $\tilde{a}_{\phi}$~\cite{note-tilde} represents
the overall strength of the heterogeneity.

For times $\tau_{\alpha} \alt t \ll \tau_{\rm ex}$, most slow (fast)
regions will stay slow (fast), so that each of their local clock
drifts $\int_0^t \delta \gamma(\vec{r},t)$ will grow linearly with
time. Thus for $\tau_{\alpha} \alt t \ll \tau_{\rm ex}$, $\tilde{\chi}^{\phi}_4(t)
\propto \langle (\text{local clock drift})^2 \rangle \propto
t^2$. (Fig.~\ref{fig:chi-phi4} shows that this regime can be found for
HARD but not for KALJ.)  Equivalently, for $0 < t' < t \ll \tau_{\rm
  ex}$, $\tilde{\chi}_2^{\phi}(t') \approx \tilde{\chi}_{2,
  M}$~\cite{note-tilde}. Thus 
Eq.~(\ref{eq:chi4-ineq-a-phi-chi4b0}) becomes an approximate equality,
$\chi_4(t) \approx \chi_{4, \rm fh}(t)$. We refer to this as the {\em
  frozen heterogeneity approximation\/}. In this approximation, the
time dependence of the dynamic susceptibility $\chi_4(t)$ is given by
an explicit expression in terms of independently measured quantities
-- $C(t)$ and $\langle (\delta N)^2 \rangle$ -- plus a single
numerical constant, $\tilde{a}_{\phi}$.

By contrast, for $t \agt \tau_{\rm ex}$ the inequality in
Eq.~(\ref{eq:chi4-ineq-a-phi-chi4b0}) becomes strict. For $t \gg
\tau_{\rm ex}$, a given region will pass through periods in which it
is slow and periods in which it is fast, so local drifts will
alternate between positive and negative signs, and
$\tilde{\chi}^{\phi}_4(t) \propto \langle (\text{local clock drift})^2
\rangle \ll \text{(const)} \; t^2$. In particular, if $\tilde{\chi}_2^{\phi}(t)$ decays
like $t^{-1-\delta} \; (\delta > 0)$ or faster, then
$\tilde{\chi}^{\phi}_4(t) \propto t \ll t^2$.

Fig.~\ref{fig:compare-model-simulations} shows a comparison of our
model's predictions for $\chi_4(t)$ with numerical
results~\cite{OhioSupercomputerCenter1987} for HARD
and KALJ. To explore the effects of the exchange time on the results,
we choose the simple parametrization $\tilde{\chi}_2^{\phi}(t) =
\tilde{\chi}_2^{\phi}(0) \, {\rm e}^{-t/\tilde{\tau}_{\rm ex}}$, which
by Eq.~(\ref{eq:tilde-chi4-phi-def}) gives
\begin{equation}
   \tilde{\chi}_4^{\phi}(t) = \tilde{a}_{\phi} \{2 \tilde{\tau}_{\rm ex}^2
      [\exp(-t/{\tilde{\tau}_{\rm ex}}) - 1 + t/\tilde{\tau}_{\rm
          ex}]\}.
   \label{eq:tilde-chi4-exp-case}
\end{equation}
There is a narrow band of possible results of the model between the
cases of frozen heterogeneity [$\tilde{\tau}_{\rm ex} = \infty,
  \chi_4(t) = \chi_{4, \rm fh}(t)$] and of $\tilde{\tau}_{\rm ex} =
\tau_{\alpha}$. The numerical results fall within that range for $t
\agt \tau_{\alpha}$ for HARD at $\varphi=0.55, 0.57$ and for $t \agt
0.1 \tau_{\alpha}$ for all other
cases~\cite{note-details-supplemental}. For HARD they approach the frozen
heterogeneity upper bound $\chi_{4, \rm fh}(t)$
as the packing fraction increases. For KALJ, they are closer to the
$\tau_{\rm ex} = \tau_{\alpha}$ results. In fact, remarkable agreement
can be obtained if the form in Eq.~(\ref{eq:tilde-chi4-exp-case}) is
used, with $\tilde{\tau}_{\rm ex}$ as a fitting parameter. (For KALJ
at $T=0.50$, the data are too noisy to judge on goodness of fit.)

\begin{figure}[ht!]
  \centering
  \includegraphics[width=.9\columnwidth] {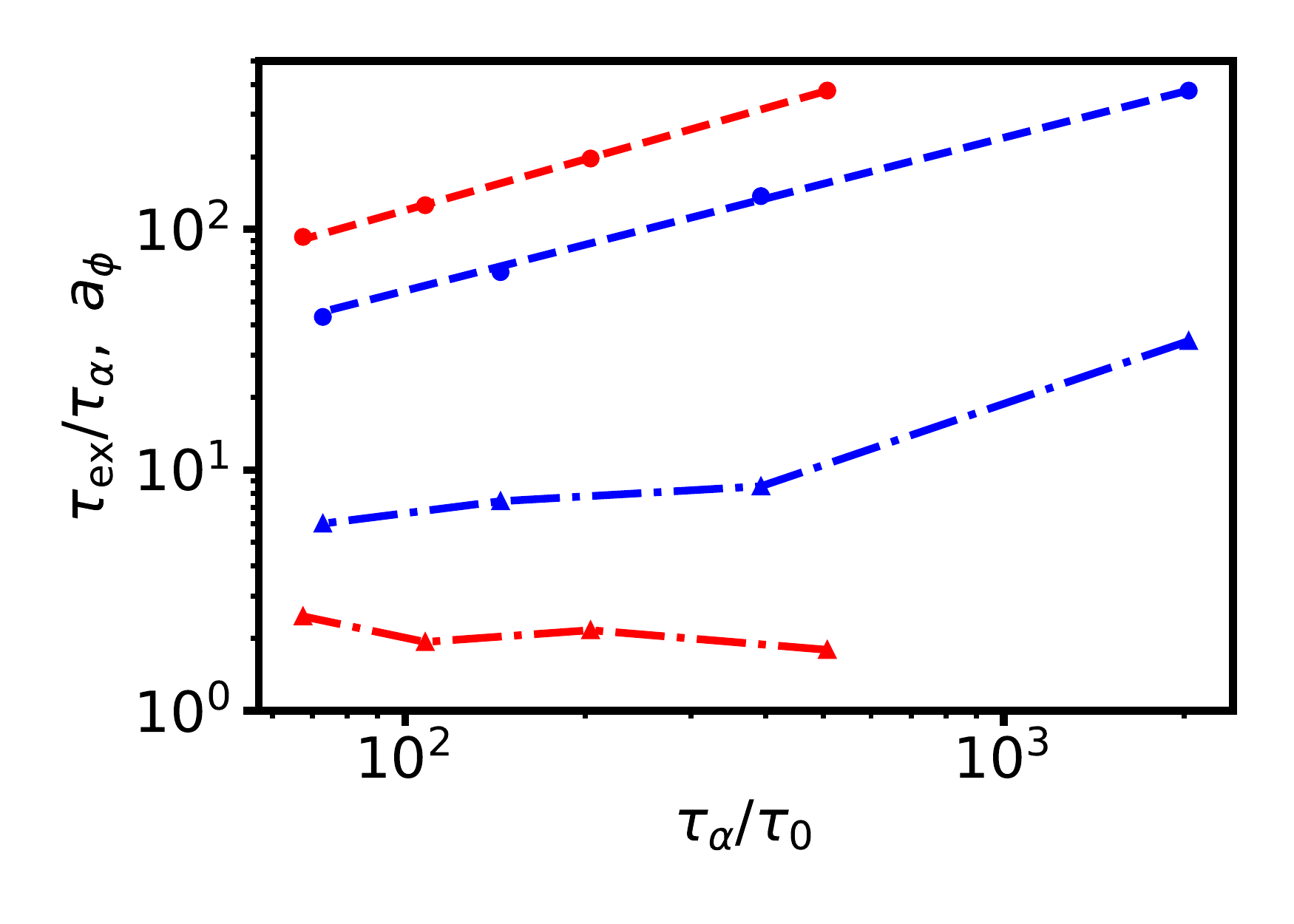}
  \caption{Ratio $Q = \tau_{\rm ex}/\tau_{\alpha}$ (triangles) and
    heterogeneity strength parameter $a_{\phi}$ (circles), vs 
    rescaled relaxation time
    $\tau_{\alpha}/\tau_0$~(\cite{Flenner2014}), for HARD (blue) at
    $\varphi= 0.55,0.56,0.57,0.58$ and KALJ (red) at $T= 0.70, 0.65,
    0.60, 0.55$. The dashed lines are power law fits for
    $a_{\phi}$.}\label{fig:a_phi}
\end{figure}

\paragraph{Results: Exchange time $\tau_{\rm ex}$}
Fig.~\ref{fig:a_phi} shows initial estimates of the memory parameter
$Q \equiv \tau_{\rm ex}/\tau_{\alpha}$~\cite{Richert2015} and the
heterogeneity strength parameter $a_\phi$, obtained under the
assumption that $\chi_2^{\phi}(t) = \chi_2^{\phi}(0) \, {\rm
  e}^{-t/\tau_{\rm ex}}$, which corresponds to $\chi_4(t) =
\dot{C}^2(t) \, a_{\phi} \, \{2 {\tau}_{\rm ex}^2
    [\exp(-t/{{\tau}_{\rm ex}}) - 1 + t/{\tau}_{\rm ex}]\} +
    \chi_{4,b}(t)$~\cite{note-details-supplemental}.
For KALJ $Q \sim 2$ and it seems not to vary strongly with
$\tau_{\alpha}$~\cite{Kim2013}, but for HARD, $Q$ increases from $\sim
5$ to $\gtrsim 30$ as the glass transition is approached. For both
models, $a_{\phi} \sim \tau_\alpha^{p} $, with $p_{HARD}
\approx 0.64$ and $p_{KALJ} \approx 0.70$.

\paragraph{Summary.}
We have introduced a simple phenomenological model for dynamic
heterogeneity in glass-forming materials. This model translates into
quantitative predictions the intuitive description of dynamic
heterogeneities as local fluctuations in the relaxation rate, and
additionally takes into account contributions due to local particle
density fluctuations and to single-particle, non-collective
behavior.

The model provides expressions for computing four-point functions -
such as Eqs.~(\ref{eq:chi4-chi4tilde-chi40})-(\ref{eq:tilde-chi2-def})
- requiring only one- and two-point quantities - like $\rho$,
$S(\vec{q})$, $C(t)$, and the relaxation rate two-point correlation
function $s(\vec{q},t)$ - that are much easier to interpret than
four-point functions.  Crucially, Eq.~(\ref{eq:chi4-chi4tilde-chi40}),
shows that $\chi_4(t)$ having a peak at $t \sim \tau_{\alpha}$ is due
to $\dot{C}^2(t)$ having its characteristic time at $t \sim
\tau_{\alpha}$, even though $\tilde{\chi}^{\phi}_4(t)$ is the factor
that most directly encodes the time dependence of relaxation rate
fluctuations. This explains why the lifetime of the dynamic
heterogeneities has a weak effect on the location $t = t_4 \sim
\tau_{\alpha}$ of the peak of $\chi_4(t)$.

Given a single parameter $\tilde{a}_{\phi}$ encoding the overall
strength of the relaxation rate fluctuations, the model predicts an
upper bound $\chi_{4, \rm fh}(t)$ for the dynamic susceptibility
$\chi_4(t)$, corresponding to the limit of frozen heterogeneity,
$\tilde{\tau}_{\rm ex} = \infty$.  If, additionally, it is assumed
that the two-point susceptibility decays exponentially,
$\tilde{\chi}_2^{\phi}(t) = \tilde{\chi}_2^{\phi}(0) \, {\rm
  e}^{-t/\tilde{\tau}_{\rm ex}}$, very good agreement is obtained with
numerical data by fitting with the two parameters $\tilde{a}_{\phi},
\tilde{\tau}_{\rm ex}$.
The dependence of $\chi_4(t)$ on $\tilde{\tau}_{\rm ex}$ is relatively
weak, and all numerical data fall in the narrow interval between the
$\tilde{\tau}_{\rm ex} = \tau_{\alpha}$ prediction and the frozen
heterogeneity $\tilde{\tau}_{\rm ex} = \infty, \chi_4(t) = \chi_{4,
  \rm fh}(t)$ prediction.

Information about the time evolution of the heterogeneities can be
found most clearly in the time evolution of $\chi_4(t)$ {\em after\/}
its peak. Roughly speaking, ${\tau}_{\rm ex} = \infty$ makes the
heterogeneities maximally persistent, which maximizes the strength of
the heterogeneity, i.e. $ \chi_4(t) \approx \chi_{4, \rm fh}(t)$, but
the shorter ${\tau}_{\rm ex}$ is, the faster the heterogeneous
relaxation rates return to the mean, and the more $\chi_4(t)$ differs
from $\chi_{4, \rm fh}(t)$ for times $t \agt \tau_{\rm ex}$.
Estimating $Q = \tau_{\rm ex}/\tau_{\alpha}$ from $\chi_4(t)$ shows
that $Q_{\rm KALJ} \sim 2$ independently of
$\tau_{\alpha}$~\cite{Kim2013}, but $Q_{\rm HARD}$ grows strongly with
$\tau_{\alpha}$. (We believe this to be the first measurement of
$Q$ or $\tau_{\rm ex}$ for a hard sphere model.) These results are consistent
with the fact that for HARD at higher packing fraction (but not for
KALJ), there is a time regime $\tau_{\alpha} \alt t \ll \tau_{\rm ex}$
where the frozen heterogeneity approximation holds,
i.e. $\tilde{\chi}^{\phi}_4(t) \sim t^2$ (Fig.~\ref{fig:chi-phi4}) and
$\chi_4(t) \approx \chi_{4, \rm fh}(t)$
(Fig.~\ref{fig:compare-model-simulations}).

Finally, the model introduces $s(\vec{q},t)$, which quantifies more
directly the spatial and temporal correlations of the local relaxation
rate.
Results for this quantity, plus discussions of non-TTI dynamics, more
general overlap functions $w_n(t)$, four-point functions with general
time arguments, and connections to experimentally measured
correlators~\cite{Ediger2000,Sillescu1999,Richert2015, Paeng2015},
will be reported elsewhere~\cite{Pandit2023,Castillo2023}.

\paragraph{Acknowledgements} We thank Elijah Flenner for providing the
numerical simulation data that was analyzed in this work.
R.~K.~P.~acknowledges the Ohio University Condensed Matter and Surface
Sciences (CMSS) program for support through a studentship.

\bibliographystyle{apsrev4-1}
\bibliography{phir-chi4} 

\begin{thebibliography}{8}
\bibitem [{not()}]{note-extrapolation-aphi}%
  \BibitemOpen
  \href@noop {} {}\bibinfo {note} {~ In particular the estimation of the limit
  $t \to 0$ for $\tilde{a}_{\phi}$ is an extrapolation where data are excluded
  for shorter times.}\BibitemShut {Stop}%
\end{thebibliography}

\end{document}


\title{Supplemental Material for:\\
  A simple model for dynamic heterogeneity in glass-forming liquids}
\author{Rajib K Pandit\footnote{Present address: Exact Sciences,
    Redwood City, 94063 CA, USA }} 
\author{Horacio E. Castillo}
\email[]{castillh@ohio.edu}
\affiliation{
  Department of Physics and Astronomy and Nanoscale
  and Quantum Phenomena Institute, Ohio University, Athens, Ohio 45701,
  USA}

\date{\today}
 
\maketitle

\subparagraph{Numerical data and data analysis}

We analyze numerical simulation data for two 3D equilibrium
glass-forming liquids. The first is a set of Monte Carlo simulations
of a 50:50 binary mixture of $N=80000$ hard-spheres (HARD), with
diameters $d$ and $1.4d$, at packing fractions $\varphi =
0.50,0.52,0.55,0.56,0.57$, and $0.58$~\cite{Flenner2011}. The second
is a set of Newtonian dynamics simulations of an $N=27000$
Kob-Andersen Lennard-Jones (KALJ)~\cite{Kob1994, Kob1995, Kob1995a}
system, at a density $\rho=N/V=1.2$, with temperatures $T= 0.50, 0.55,
0.60, 0.65, 0.70$, and $0.80$~\cite{Flenner2014}. More details on the
simulations can be found in Ref.~\cite{Flenner2011} for HARD and in
Ref.~\cite{Flenner2014} for KALJ.

To obtain $\chi_4(t)$, we
fitted~\cite{OhioSupercomputerCenter1987} 
$S_4(\vec{q},t)$ with the form~\cite{Pandit2022}
\begin{equation}
  \label{eq:modified-oz}
  S_4(\vec{q},t)=\frac{\chi_4^{\rm cr}(t)}{1+ [\xi_4^{\rm cr}(t)]^2 q^2
   + [c(t)]^2 q^4} + \chi_{4,b}(t), 
\end{equation} 
where $\chi_4^{\rm cr}(t)$ is the collective relaxation part of the
dynamic susceptibility, $\xi_4^{\rm cr}(t)$ is the four-point dynamic
correlation length, and $\chi_4(t) = \chi_4^{\rm cr}(t) +
\chi_{4,b}(t)$.
For $t$ long enough that Eq.~(12) in the main text is expected
to be accurate~\cite{note-extrapolation-aphi}, we
extract~\cite{OhioSupercomputerCenter1987} 
\begin{equation}
  \hspace{-0.3cm}\tilde{\chi}^{\phi}_4(t) = 
  [\chi_4(t) - \chi_{4,b}^{(0)}(t)]
  \dot{C}^{-2}(t), 
  \;\; \text{and} \;\;
  \tilde{a}_{\phi} = \lim_{t \to 0} \; t^{-2} \!\tilde{\chi}^{\phi}_4(t).
  \label{eq:aphi-determ}
\end{equation}
%

\subparagraph{Range of validity of the approximations shown in Fig.~3 in
  the main text}

A likely explanation for the fact that the model starts to represent
well the data at earlier $t/\tau_{\alpha}$ fractions in some cases is
that for the glassier cases, there is a stronger separation between
the timescales of the first and second step in the two-step
relaxation, which makes Eq.~(12) in the main text - and consequently
Eqs.~(13) and (17) in the main text - accurate over a wider range of
timescales, and in particular even for times shorter than
$\tau_{\alpha}$.

\renewcommand{\thefigure}{SM-1}
\begin{figure}[hb!]
  \centering
  \includegraphics[width=.9\columnwidth] {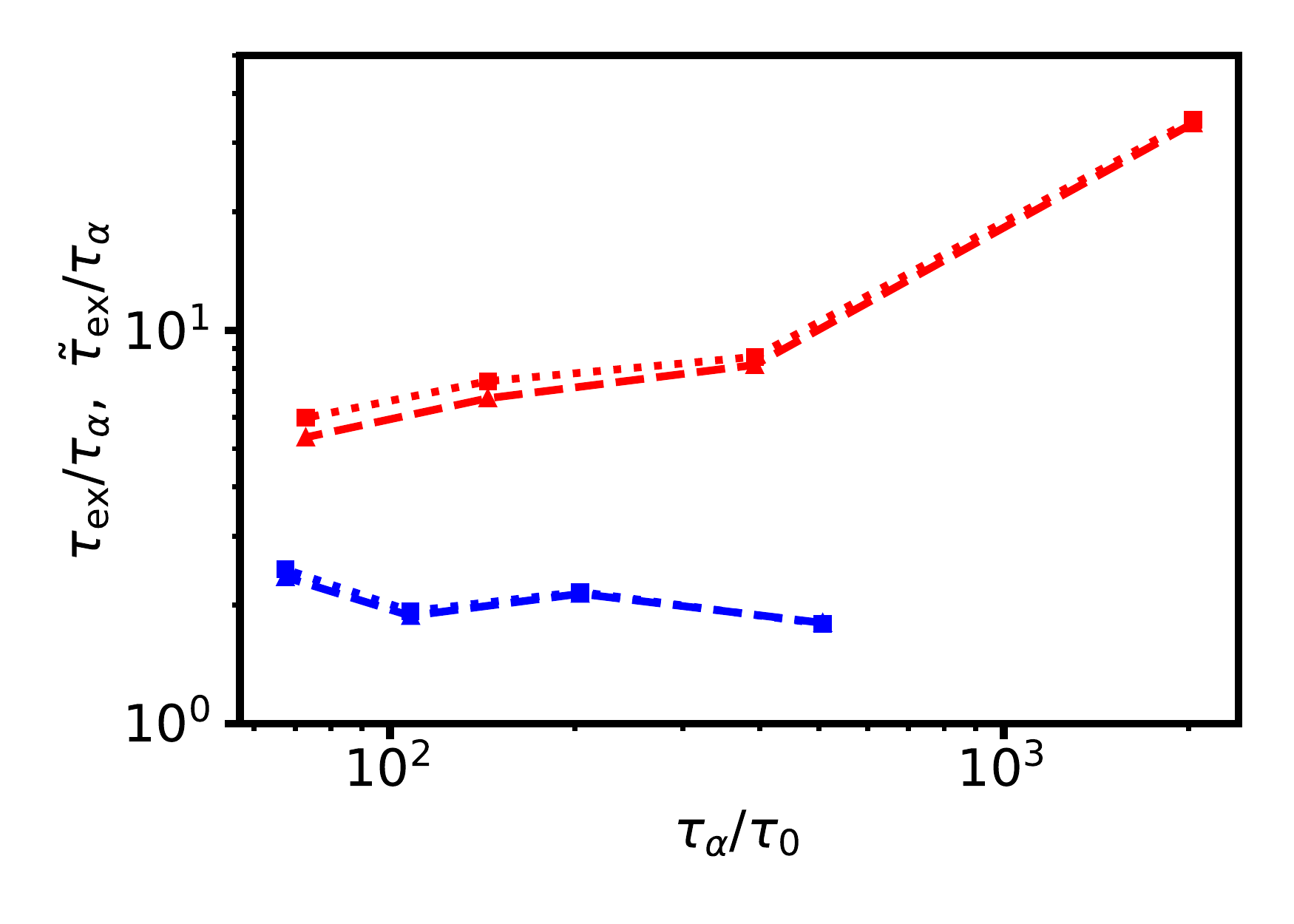}
 \caption{Ratios $\tau_{\rm ex}/\tau_{\alpha}$ (triangles) and
   $\tilde{\tau}_{\rm ex}/\tau_{\alpha}$ (squares), as functions of
   the rescaled relaxation time
   $\tau_{\alpha}/\tau_0$~(\cite{Flenner2014}) for HARD (blue) at
   $\varphi= 0.55,0.56,0.57,0.58$ and KALJ (red) at $T= 0.70, 0.65,
   0.60, 0.55$. }\label{fig:SM-1}
\end{figure}

\subparagraph{$S_4^{\rm mc}({\vec{q}},t)$, and the relations between
  $\tilde{\chi}_2^{\phi}(t)$ and ${\chi}_2^{\phi}(t)$, and between
  $\tilde{\tau}_{\rm ex}$ and ${\tau}_{\rm ex}$.}

$\tilde{\tau}_{\rm ex}$ characterizes the decay with time of
$\tilde{\chi}_2^{\phi}(t)$ [Eq.~(16)], which incorporates a correction
to ${\chi}_2^{\phi}(t)$ [Eq.~(10)] due to the interplay of density and
relaxation rate fluctuations. It differs slightly from $\tau_{\rm
  ex}$, which characterizes only the relaxation rate fluctuations
probed by ${\chi}_2^{\phi}(t)$, as shown in Fig.~\ref{fig:SM-1}.

In the case of $\tilde{\tau}_{\rm ex}$, combining Eqs.~(13) and (19)
in the main text leads to
\begin{equation}
   \chi_4(t) = \dot{C}^2(t) \, \tilde{a}_{\phi} \,
   \{2 \tilde{\tau}_{\rm ex}^2 [\exp(-t/{\tilde{\tau}_{\rm ex}}) - 1
     + t/\tilde{\tau}_{\rm ex}]\} + \chi_{4,b}^{(0)}(t).
   \label{eq:chi4-aphi-tau-tilde-chi40}
\end{equation}
Here the background value $\chi_{4,b}^{(0)}(t)$ is separately
determined from $C(t)$ and $S(q)$ by using Eq.~(14) in the main text.

In the case of ${\tau}_{\rm ex}$, we repeat the derivation that leads
to Eq.~(13) in the main text, but we collect terms in a slightly
different way. Thus we obtain 
\begin{eqnarray}
   && \hspace{-1.0cm}
   \chi_4(t) \equiv \lim_{q \to 0} S_{4}({\vec{q}},t)  = 
   \dot{C}^2(t) \; {\chi}^{\phi}_4(t)
   + \chi_{4,b}(t), \quad \text{with}
   \label{eq:chi4-chi4-chi4b}
   \\
   && \hspace{-1.0cm}
   {\chi}^{\phi}_4(t) \equiv 
   \rho \tau^2
   \!\! \textstyle\int_{0}^{t} \! dt'' \!\!
   \textstyle\int_{0}^{t} \! dt'
   {\chi}_2^{\phi}(t''-t'),
   \label{eq:chi4-phi-def}
   \\
   && \hspace{-1.0cm}
   \chi_{4,b}(t) 
   \approx 
   \chi_{4,b}^{(0)}(t) + 
   \lim_{q \to 0} S_4^{\rm mc}({\vec{q}},t), 
   \label{eq:chi4b-def}
   \quad \text{and}
   \\
   && \hspace{-1.0cm}
   S_4^{\rm mc}({\vec{q}},t) =
   \dot{C}^2(t) 
   \tau^2
   \!\! \textstyle\int_{0}^{t} \! dt'' \!\!
   \textstyle\int_{0}^{t} \! dt' \!\!
   \left\{[S_c(\cdot)-1]*s(\cdot,t''-t')\right\}\!(\vec{q}). 
\end{eqnarray}
Here ${\chi}_2^{\phi}(t)$ and $s(\vec{q},t)$ are defined in Eqs.~(10)
and (11) in the main text, $S_c({\vec{q}}) \equiv [S({\vec{q}}) - \rho
  \delta_{\vec{q}, \vec{0}}]$ is the connected part of the static
structure factor and $f*g$ denotes a convolution. We notice that
$\chi_{4,b}(t)$ is the same quantity that was defined in
Ref.~\cite{Pandit2022}. By assuming that $\chi_2^{\phi}(t) =
\chi_2^{\phi}(0) \, {\rm e}^{-t/\tau_{\rm ex}}$, the equations above
lead to
\begin{equation}
   \chi_4(t) = \dot{C}^2(t) \, a_{\phi} \, \{2 {\tau}_{\rm ex}^2
       [\exp(-t/{{\tau}_{\rm ex}}) - 1 + t/{\tau}_{\rm 
    ex}]\} + \chi_{4,b}(t).
   \label{eq:chi4-aphi-tau-chi4b}
\end{equation}

From the numerical point of view, we see that the difference between
$\tilde{\chi}_2^{\phi}(t)$ (or $\tilde{\tau}_{\rm ex}$) and
$\chi_2^{\phi}(t)$ (or $\tau_{\rm ex}$) comes from the difference between
the $\chi_{4,b}^{(0)}(t)$ background in the first case and the
$\chi_{4,b}(t)$ background in the second. From the point of view of
{\em predicting\/} the behavior of $\chi_4(t)$, the former is
preferable because $\chi_{4,b}^{(0)}(t)$ is completely determined by
the global average quantities $C(t)$ and $S(q)$, i.e.~it does not
require any prior information regarding dynamic heterogeneity. For
this reason it is used in the analysis shown in Figs.~2 and 3 in the
main text. By contrast, from the point of view of estimating the
lifetime of heterogeneities, $\tau_{\rm ex}$ is preferable to
$\tilde{\tau}_{\rm ex}$, because it is the same quantity defined in
Eq.~(10) in the main text, which does not include any extra
contributions associated with density fluctuations, and for this
reason $\tau_{\rm ex}$ and $a_{\phi}$ are the quantities that are
shown in Fig.~4 in the main text.

In practice, as pointed out in Ref.~\cite{Pandit2022}, the difference
between $\chi_{4,b}(t)$ and $\chi_{4,b}^{(0)}(t)$ is very small. This
implies that the difference between $\tilde{\chi}_2^{\phi}(t)$ and
$\chi_2^{\phi}(t)$, and the difference between $\tilde{\tau}_{\rm ex}$
and $\tau_{\rm ex}$ are both small. In the case of the characteristic
times, this is confirmed by the results shown in  Fig.~\ref{fig:SM-1}.

\bibliographystyle{apsrev4-1}
\bibliography{phir-chi4_SM}